\documentclass[pra,twocolumn,superscriptaddress,amsmath,amssymb,aps,10pt]{revtex4-1}

\usepackage{graphicx}
\usepackage{dcolumn}
\usepackage{bm}
\usepackage{braket}
\usepackage{hyperref}
\hypersetup{
    colorlinks=true,        
    linkcolor=blue,          
    citecolor=blue,        
    filecolor=magenta,      
    urlcolor=blue           
}

\renewcommand{\ij}{\langle ij\rangle}
\newcommand{\ua}{\uparrow}
\newcommand{\da}{\downarrow}
\newcommand{\dg}{\dagger}
\newcommand{\pdg}{{\phantom{\dagger}}}

\newcommand{\sg}{\sigma}
\newcommand{\gm}{\gamma}

\usepackage{color} 				
\usepackage[normalem]{ulem} 	

\begin{document}

\title{Suppression and revival of long-range ferromagnetic order in the multiorbital Fermi-Hubbard model}

\author{Andrii Sotnikov}
\email{sotnikov@ifp.tuwien.ac.at}
\affiliation{Institute of Solid State Physics, TU Wien, Wiedner Hauptstra{\ss}e 8, 1020 Vienna, Austria}
\affiliation{Akhiezer Institute for Theoretical Physics, NSC KIPT, Akademichna 1, 61108 Kharkiv, Ukraine}
\author{Agnieszka Cichy}
\affiliation{Faculty of Physics, Adam Mickiewicz University, Umultowska 85, 61-614 Poznan, Poland}
\affiliation{Institut f\"{u}r Physik, Johannes Gutenberg-Universit\"{a}t Mainz, Staudingerweg 7, D-55099 Mainz, Germany}
\author{Jan Kune\v{s}}
\affiliation{Institute of Solid State Physics, TU Wien, Wiedner Hauptstra{\ss}e 8, 1020 Vienna, Austria}

\date{\today}

\begin{abstract}
By means of dynamical mean-field theory allowing for complete account of SU(2) rotational symmetry of interactions between spin-1/2 particles, we observe a strong effect of suppression of ferromagnetic order in the multiorbital Fermi-Hubbard model in comparison with a widely used restriction to density-density interactions. In the case of orbital degeneracy, we show that the suppression effect is the strongest in the two-orbital model (with effective spin $S_{\rm eff}=1$) and significantly decreases when considering three orbitals ($S_{\rm eff}=3/2$), thus magnetic ordering can effectively revive for the same range of parameters, in agreement with arguments based on vanishing of quantum fluctuations in the limit of classical spins ($S_{\rm eff}\to\infty$). We analyze a connection to the double-exchange model and observe high importance of spin-flip processes there as well.
\end{abstract}

\maketitle

\section{Introduction}

Symmetry, its discrete or continuous nature, and its explicit or spontaneous breaking play a crucial role in physics. In condensed-matter theory, Heisenberg and Ising models are distinctive examples of systems possessing continuous and discrete spin symmetry, respectively. While the spontaneous breaking of spin symmetry plays a central role in observations of phase transitions and emerging gapless collective modes, its explicit analog is less ``charming'' and usually originates from the external source fields, sample imperfections, or simplifications required to proceed with the corresponding theoretical description.

Long-range ferromagnetic (FM) order is prominent realization of spontaneously broken symmetry responsible for many important physical phenomena, e.g., the colossal magnetoresistance effect in manganites~\cite{Moritomo1996}. In a search of a prototypical lattice system of itinerant interacting particles supporting the FM ground state, a single-band Hubbard model can be suggested as the simplest one. However, as emphasized and studied in detail in Refs.~\cite{Stollhoff90,Kollar1996PRB}, this is {\it not a generic} model of ferromagnetism, since without lattice loop structures or additional nearest-neighbor ``biasing'' interactions the corresponding FM ground state can only emerge in the Nagaoka's limit ($U=\infty$). The next by simplicity, the two-orbital Hubbard model, provides a minimal number of necessary ingredients (in particular, a local nonzero Hund's coupling) to support FM ordering.

Nowadays, dynamical mean-field theory (DMFT) \cite{Georges1996RMP} is a powerful nonperturbative theoretical approach to describe physics of strongly correlated materials including transitions between different thermodynamic phases. In a number of previous DMFT studies analyzing FM instability in the two-band Hubbard model the computational procedure was restricted to the density-density interactions~\cite{Held1998EPJ, Hoshino2015PRL, Hoshino2016PRB, Cichy2016PRA, sotnikov17prb}. Although this simplification substantially reduces the computational cost, it affects the physics of the model, especially close to the critical regime (see, e.g., Ref.~\cite{hausoel17} for a recent material-oriented analysis). Therefore, a significant effort has been made to account for the full rotational symmetry of two-particle interactions \cite{Kubo98, Werner2006PRB, Sakai07, Kubo09, Lauchli2009PRB, chan09, peters10, Pruschke11, Antipov2012PRB, Parragh2012PRB, Golubeva2017PRB}.

In parallel to the progress of DMFT and other theoretical approaches, ultracold atoms in optical lattices~\cite{Blo2008RMP} have become a universal and very accurate experimental tool for gaining new insights in a rich family of the Hubbard-type models. In these systems, a degree of freedom associated with the electron spin can be attributed to atoms in different internal states (i.e., the {\it pseudospin} concept is widely applied). This results in a large tunability --- at will, the spin symmetry can be explicitly broken or restored with a high accuracy by a proper experimental setting \cite{Tai2010PRL,Kra2012Nat}. In particular, recent experiments with $^{173}$Yb and $^{87}$Sr atoms~\cite{Scazza2014NP, Zhang14S, Riegger18} show the capability to realize two-orbital Hubbard models with the FM-type Hund's coupling and SU($N$)-symmetric interactions of pseudospin flavors.

The purpose of the present paper is to analyze the effect of interaction parametrization --- with the SU(2) spin-rotational symmetry or lower --- on FM ordering in the multiorbital Hubbard models. Starting with degenerate orbitals we introduce a crystal field in order to link
the metallic Stoner-type FM phase of the Hubbard model with overlapping bands to the FM phase of the double-exchange (DE, or the Kondo-lattice) model. The latter has a long history~\cite{Zener51} and it is widely applied to describe FM ordering and related effects in manganites. The model studies by means of DMFT have already uncovered important effects of electronic correlations~\cite{held00,imai00} and spin fluctuations~\cite{nagai00} on FM ordering in the DE regime.

The paper is organized as follows.
In Sec.~\ref{sec.2} we introduce the theoretical model and specify relevant details of the applied numerical approach. We analyze the FM instability starting from the case of degenerate orbitals in Sec.~\ref{sec.3A} and proceed further with introducing the chemical potential imbalance (i.e., a nonzero crystal field) in Sec.~\ref{sec.3B}. The main results are summarized in Sec.~\ref{sec.4}.

\section{Model and Method}\label{sec.2}
We consider the Fermi--Hubbard model with multiple ($m=2,3$) orbitals described by the Hamiltonian
\begin{eqnarray}
{\cal H} &=& \sum_{\bf k\sg}\sum_{\gm=1}^m (\epsilon_{\bf k\gm}+\mu_\gm) c^\dg_{\bf k\gm\sg} c^\pdg_{\bf k\gm \sg}
+ U \sum_{i\gm} n_{i\gm\ua} n_{i\gm\da}
\nonumber \\
&&+ U'\sum_{i\sg,\gm<\gm'} n_{i\gm\sg} n_{i\gm'\bar \sg}
+ (U'-J) \sum_{i\sg,\gm<\gm'} n_{i\gm\sg} n_{i\gm'\sg}
\nonumber \\
&&+ \alpha J \sum_{i,\gm<\gm'} (
 c^\dg_{i\gm\ua}c^\dg_{i\gm'\da}  c^\pdg_{i\gm\da} c^\pdg_{i\gm'\ua}+{\rm H.c.} )
\nonumber \\
&&+\alpha'J \sum_{i,\gm<\gm'}(c^\dg_{i\gm\ua} c^\dg_{i\gm\da} c^\pdg_{i\gm'\da} c^\pdg_{i\gm'\ua}
+{\rm H.c.} ). 
\label{hamilt}
\end{eqnarray}
The first term includes free-particle energies~$\epsilon_{\bf k\gm}$, chemical potentials~$\mu_\gm$, and fermionic creation (annihilation) operators $c^\dg_{\bf k\gm\sg}$ ($c_{\bf k\gm \sg}$) of electrons on the orbital~$\gm$ with the spin~$\sigma=\ua,\da$ (and its opposite $\bar \sg=\da,\ua$) and quasimomentum~{\bf k}.
$U$ is the intraorbital interaction amplitude and $J$ characterizes local ($i$ is the lattice site index) ferromagnetic ($J>0$) Hund's coupling.
Here and below, we use the parametrization $U'=U-2J$, which is generally valid for all electron-electron interactions that are rotationally invariant in real space. 
The coefficients $\alpha,\alpha'\in[0,1]$ can be set, in principle, independently of each other.
At $\alpha=\alpha'$, the two limiting cases with $\alpha=0$ and $\alpha=1$ correspond to the so-called density-density (Ising-type) and Slater-Kanamori (SK) parametrization of interactions.
We also restrict ourselves to all positive interaction amplitudes that results in limitations $U>0$ and $J\leq U/3$.

The Hamiltonian~\eqref{hamilt} allows for a separate analysis of spin and orbital sectors. Since the symmetry of the latter is almost irrelevant for the current study, we omit its discussion \footnote{
See Refs.~\cite{Hoshino2016PRB,Georges13} for details.} for simplicity. 
Therefore, it is sufficient to point out briefly the influence of the Ising-anisotropy parameter~$\alpha$ on the symmetry of the spin sector. In particular, at $\alpha=1$ the model becomes SU(2) symmetric with respect to rotations in spin space. Any other value of $\alpha$ ($0\leq\alpha<1$) lowers the symmetry of the spin part to the $\mathbb{Z}_2\times$U(1) group, where $\mathbb{Z}_2$ corresponds to reflections of the type $c_{\gm\ua}\to c_{\gm\da}$ and U(1) suggests the invariance of the Hamiltonian under rotations around the $z$ quantization axis, 
$(c_{\gm\ua},c_{\gm\da})\to (e^{i\phi}c_{\gm\ua},e^{-i\phi}c_{\gm\da})$.

In the DMFT analysis of magnetic ordering in the system under study, we employ two types of solvers for the auxiliary Anderson impurity problem. We use the continuous-time hybridization expansion quantum Monte Carlo (CT-HYB) impurity solver provided via the \textsc{w2dynamics} software package~\cite{w2dynamics}, which includes necessary generalizations introduced in Refs.~\cite{Werner2006PRB,Lauchli2009PRB,Parragh2012PRB}. The second option, the exact diagonalization (ED) solver is based on extensions discussed in Refs.~\cite{Sotnikov2015PRA,Golubeva2017PRB}.
The maximal number of effective bath orbitals per each orbital and spin flavor in ED is limited in the present study to $n_s=4$ (for $m=2$) and $n_s=3$ (for $m=3$)
\footnote{The limitations in ED are caused by the exponential growth of the corresponding Hilbert space with the total number of states ${\cal N}=2mn_s$. We systematically verified that the used maximal numbers of $n_s$ are sufficient for ED and CT-HYB results to fall into a good agreement of the order of the line width in figures corresponding to the regime of overlapping bands. At lower $n_s$ values or in the DE regime, the deviations become more noticeable; see also Appendix~\ref{app.A}}.
Since the used CT-HYB solver does not suffer from a sign problem~\cite{Werner2006PRB}, we use its output for the resulting diagrams and graphical dependencies and the output from the ED solver as a supplementary source supporting main observations.

For simplicity and general analysis purpose, it is sufficient to restrict below to a semicircular density of states $D(\epsilon)=(1/2\pi t^2)\sqrt{4t^2-\epsilon^2}$
and to set the hopping amplitude~$t$ as the scaling unit ($t\equiv t_m=1$) in all energy-related quantities (e.g., the bandwidth for the non-interacting system thus becomes $W=4$). Note that estimates for more realistic lattice geometries can usually be performed by a proper rescaling of quantities with respect to the coordination number~$z$ of the actual lattice ($t\to \sqrt{z}t$) \cite{Blumer2013PRB}. For example, in simple cubic (sc) lattice geometry with $z=6$, the energy-related parameter $P$ (e.g., critical temperature or interaction strength) becomes $P_{\rm{sc}}\approx\sqrt{6}P_{\rm{Bethe}}$. At weak coupling the rescaling holds only if there are no Van Hove singularities close to the Fermi level, otherwise, the magnetic ordering tendency can be enhanced. At strong coupling ($U\gg t$), the shape of the noninteracting density of states is less relevant.

The FM instability is analyzed in two ways: by direct measurements of uniform magnetizations (both with the CT-HYB and ED impurity solvers) and by analyzing uniform susceptibilities in the symmetric phase with an external magnetic field (with the ED solver, see, e.g., Ref.~\cite{Cichy2016PRA}). Due to restricting to the hybridization functions that are diagonal in orbital and spin indices, we do not have direct access in measurements of other potentially competing (e.g., canted antiferromagnetic, Ruderman-Kittel-Kasuya-Yosida coupling, spin-orbit pairing, and excitonic) instabilities.

\section{Results}\label{sec.3}

\subsection{Degenerate orbitals}\label{sec.3A}
\begin{figure}
\includegraphics[width=\linewidth]{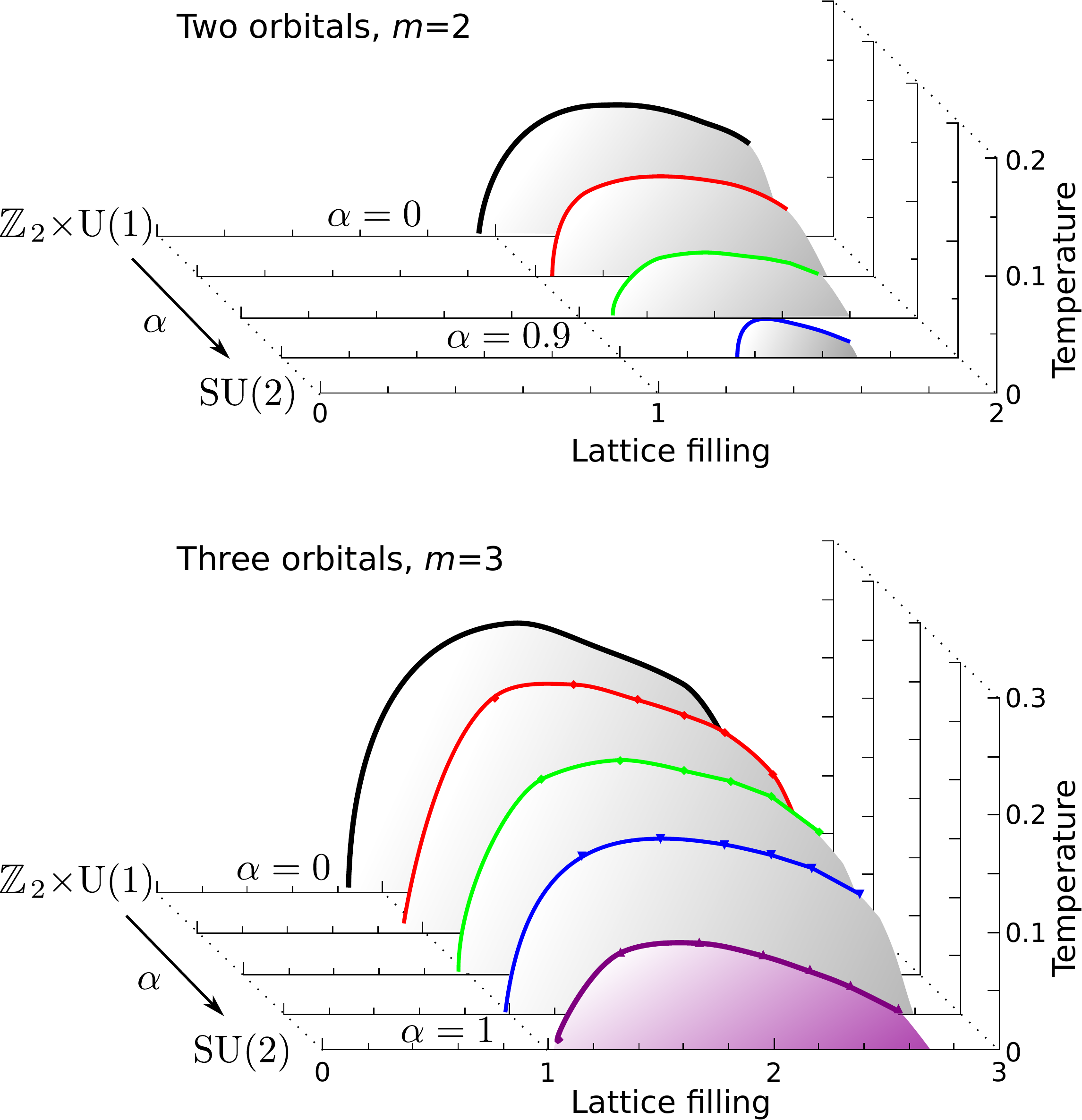}
    \caption{\label{fig1}
    Phase diagrams indicating evolution of FM phases under the change of the Ising anisotropy parameter~$\alpha$ ($\alpha=0,0.6,0.8,0.9,1$ from back to front) at $U=12$ and $J=3$.}
\end{figure}

In Fig.~\ref{fig1} we show the FM phase boundary as a function of temperature and filling for several values of the parameter $\alpha$. The band filling and interactions parameters, $U=12$ and $J=U/4$ \footnote{In particular, at $m=3$ and $n=2$ this is far away from the Mott transition, see Ref.~\cite{deMedici11}}, fall to the Hund's metal regime~\cite{yin11, Georges13} with a sizable dynamical mass enhancement.
The most striking feature is that in the case of a two-orbital system FM order completely vanishes in the SU(2)-symmetric ($\alpha=1$) limit.
This is in contrast to previous expectations (based on DMFT analysis with the density-density interactions only; see Ref.~\cite{Held1998EPJ}) that the spin-flip term has no strong effect on critical temperatures in this parameter range. In fact, it does and, as we see from comparison with the three-band model, FM ordering is sensitive to the number of active orbitals. We attribute this effect to suppression of quantum fluctuations with increasing orbital degeneracy, i.e., effective moments become more classical (see, e.g., Ref.~\cite{Fazekas1999}).

\begin{figure}
\includegraphics[width=\linewidth]{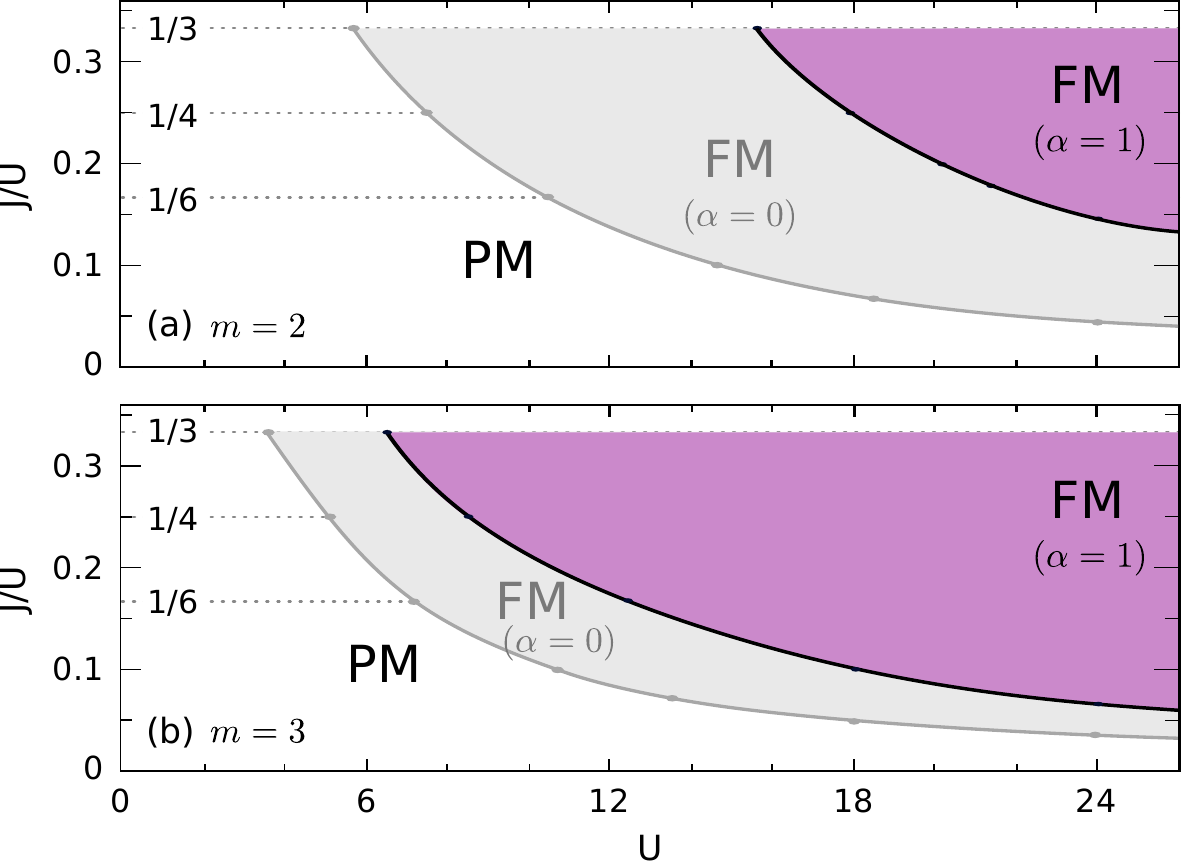}
    \caption{\label{fig2}
    $U$-$J$ phase diagrams of the degenerate two-orbital (a) and three-orbital (b) models in the low-temperature region ($T=0.025$) at
    the lattice filling $n=1.5$.
    }
\end{figure}

In Fig.~\ref{fig2}, we compare the critical interaction parameters at
fixed temperature ($T=0.025$) for the limiting cases $\alpha=0$ and $\alpha=1$.
For the two-orbital model, we observe a strong suppression of the FM phase by the spin-flip term. This effect is significantly reduced in the three-orbital model. The pronounced impact of band multiplicity agrees with the early studies on real materials~\cite{Stollhoff90}. Note that, similarly to arguments for the single-band SU(2)-symmetric Hubbard model~\cite{Kollar1996PRB}, the results confirm absence of the FM instability at $U<\infty$ in higher-symmetric (single-band) SU(2$m$)-symmetric models, corresponding to the $J\to0$ limit in Fig.~\ref{fig2}.

The form of the interaction, parameter $\alpha$, affects not only the FM phase boundaries, but also single-particle observables in the paramagnetic (PM) regime such as the effective mass and quasiparticle lifetime, relevant in Hund's metal physics~\cite{yin11,deMedici11,Georges13} . 
\begin{figure}
\includegraphics[width=\linewidth]{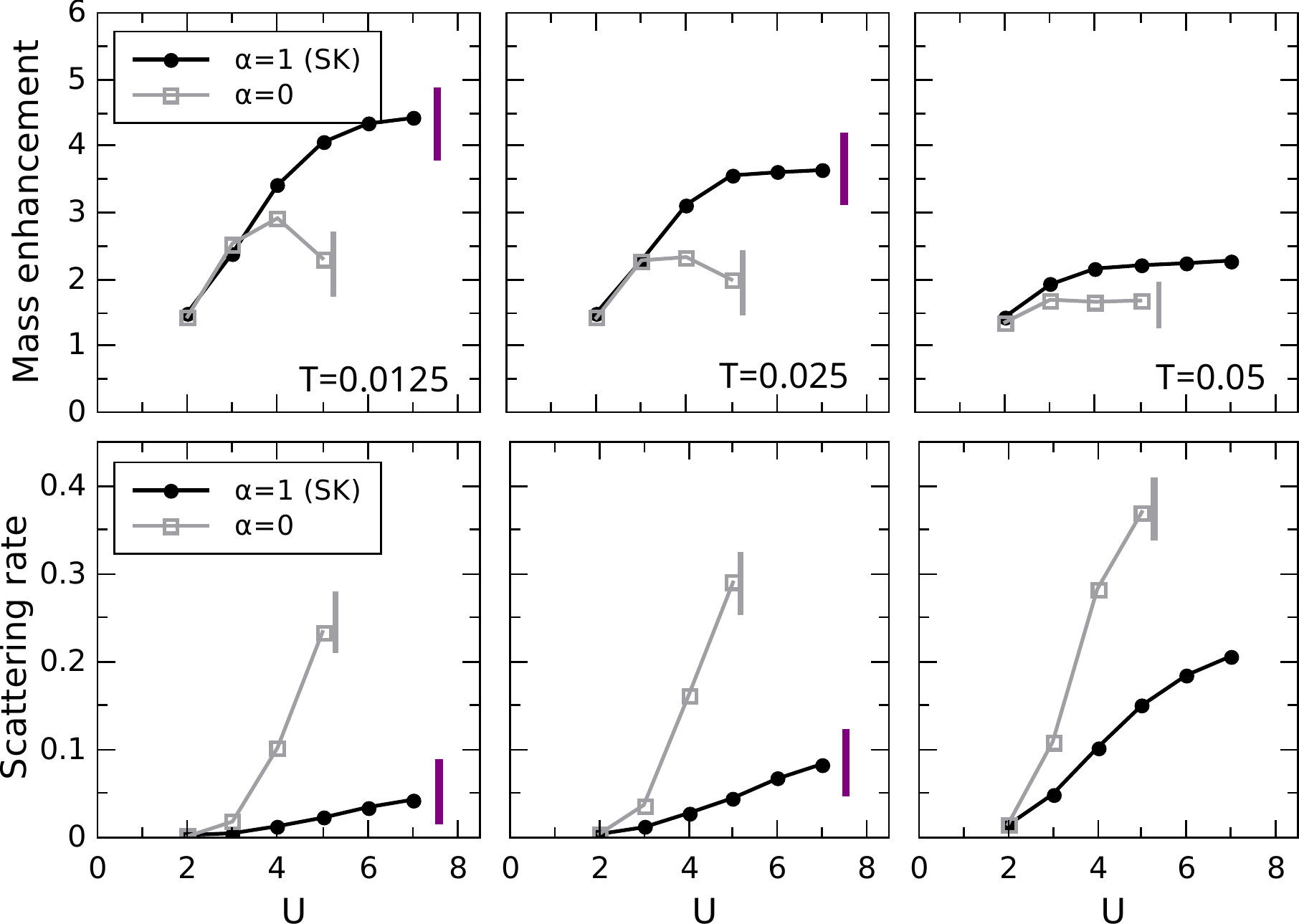}
    \caption{\label{fig_new}
    Dependencies of the mass enhancement and the quasiparticle scattering rate (inverse lifetime) on the interaction strength~$U$ for the $m=3$ model in the PM regime at $n=2$, $J=U/4$, and three different temperatures, $T=0.0125,0.025,0.05$. The vertical bars indicate the corresponding FM phase boundaries.}
\end{figure}
Following Ref.~\cite{dang15}, we use the polynomial fit to the imaginary part of the self-energy~$\Sigma(i\omega_n)$ at the six lowest Matsubara frequencies to get the quasiparticle mass enhancement $Z^{-1}=1-\left.d\operatorname{Im}[\Sigma(i\omega_n)]/d\omega_n\right|_{\omega_n\to0}$ and the quasiparticle scattering rate~$\Gamma$ (the inverse lifetime) $\Gamma Z^{-1}=-\left.\operatorname{Im}[\Sigma(i\omega_n)]\right|_{\omega_n\to0}$. As shown in Fig.~\ref{fig_new}, the difference between $\alpha=0$ and $\alpha=1$ is maximized in the vicinity of the FM($\alpha=0$) phase boundary. The results for $m=3$, $n=2$, and $\alpha=1$ agree well with the available quasiparticle weights~$Z$ of Ref.~\cite{deMedici11}. The FM phase boundary in this case matches well the one obtained for calcium ruthenate CaRuO$_3$~\cite{dang15}.

\subsection{Split orbitals}\label{sec.3B}
\begin{figure*}
\includegraphics{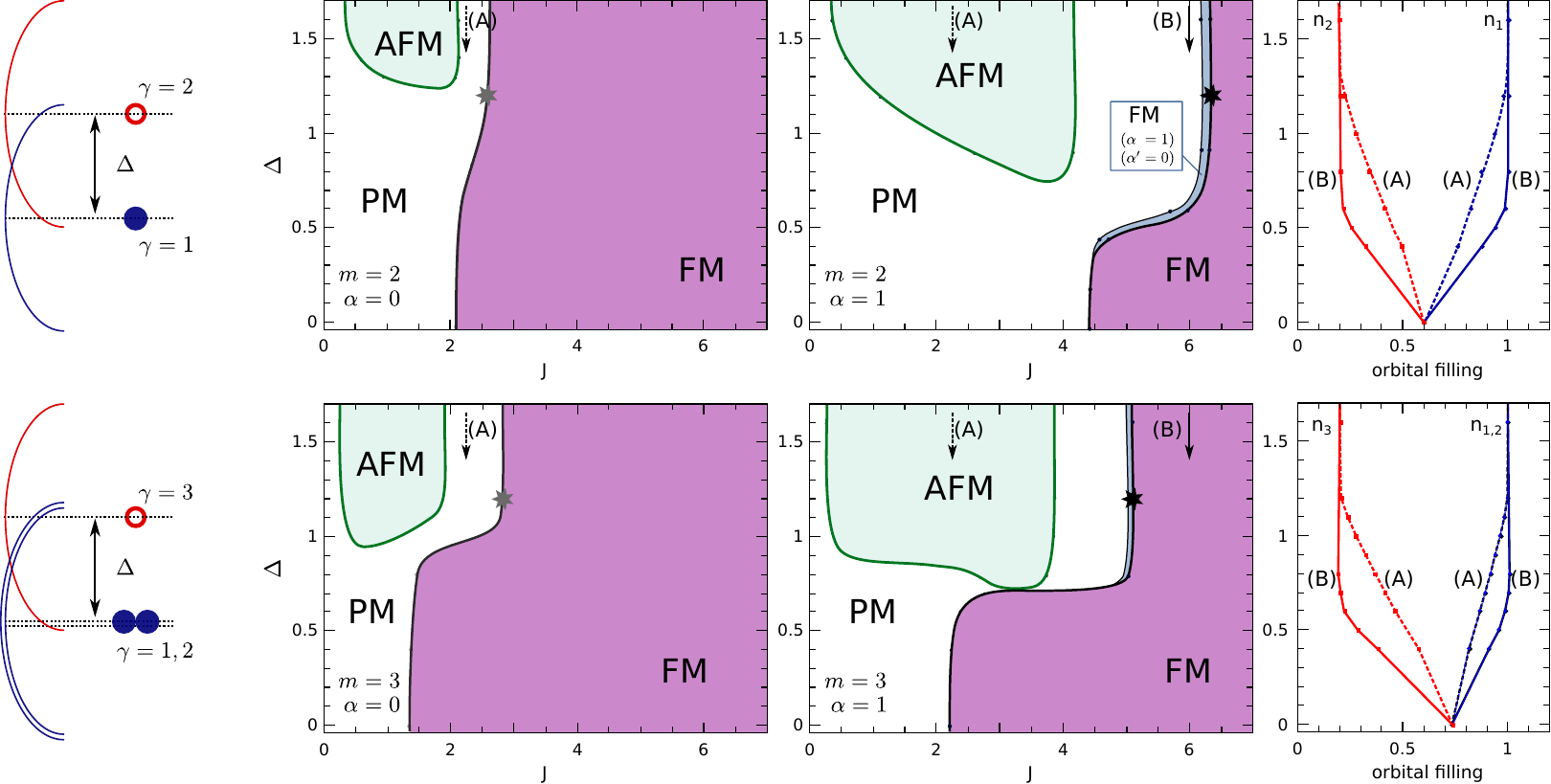}
    \caption{ \label{fig3}
    Left: Sketch of the energy states and orbital occupancies of the system with the depicted  noninteracting densities of states and the splitting~$\Delta$. Center: $J$-$\Delta$ phase diagrams of the two-orbital ($m=2$, upper row) and three-orbital ($m=3$, lower row) models at the lattice fillings $n=1.2$ and $n=2.2$, respectively. Right: the orbital occupancies $n_{\gm}$ (obtained at $\alpha=1$) along the vertical lines indicated by arrows on the corresponding diagrams at $J=2.25$ (A, dashed lines) and  $J=5.5$ (B, solid lines). The $U/J$ ratio is kept fixed,  $U=4J$, and $T=0.025$ everywhere.}
\end{figure*}
Next, we analyze the FM instability when the degeneracy is lifted by a crystal field $\Delta=\mu_{m}-\mu_{\gamma}$ ($\gamma=1,\dots,m-1$), which splits one orbital from the rest. 
The population $n=m-1+\delta$ is fixed so that for large $U$ and $\Delta>t_\gm$ (DE regime) the lower (degenerate) band becomes half filled, while setting the population of the split-off band to $\delta=0.2$. First, we keep all bandwidths equal, $W_\gm=W_m=4$, and vary $\Delta$.
Later, we will vary the bandwidth of the lower (degenerate) band for the fixed $\Delta$.

In Fig.~\ref{fig3} we show the FM phase boundaries as functions of $\Delta$ and $J$. Similar to Sec.~\ref{sec.3A}, the full spin-rotational symmetry ($\alpha=1$) favors the PM phase and the effect is less pronounced with increasing number of bands.
Two regions with weak dependencies of $J_c$ on $\Delta$ are distinguishable with a stepwise change between them. The ``step'' coincides with the onset of integer filling of the lower band. This suggests that the DE regime (at higher $\Delta$) ends abruptly at this point. At low $\Delta$, the AFM phase is not stable for any $J$. 
The pair-hopping term ($\alpha'=1$) has only a minor effect on phase boundaries (see Fig.~\ref{fig3}). The likely explanation is provided by a small number of doubly occupied orbitals in the preset Hund's coupling regime.


In the DE regime ($n_\gamma=1$ for $\gamma<m$ and $n_m=\delta$), which corresponds to the region $\Delta\gtrsim1$ and $J\gtrsim1$ in Fig.~\ref{fig3}, the system can be described by the ferromagnetic Kondo-lattice model with an additional AFM coupling between local moments \cite{deGennes60,Ishihara97,Kagan99,Yi00,Ohsawa02},
\begin{eqnarray}
	{\cal H}_{\rm eff} &=& -\sum_{\gm=1}^m t_{\gm}\sum_{\ij\sg} c^\dg_{i\gm\sg} c^{\phantom{\dg}}_{j\gm \sg} 
    +{\cal J}_{\rm A}\sum_{\ij}{\bf S}_i\cdot{\bf S}_j
    \nonumber
    \\
    &&- {J}\sum_i c^\dg_{im\sg}\boldsymbol{\tau}^\pdg_{\sg\sg'}c^\pdg_{im\sg'}\cdot{\bf S}_i,\label{extKLM}
\end{eqnarray}
where ${\bf S}_i$ are the spin operators  for the ``localized'' spins of the size $S=(m-1)/2$, ${\bf S}_i=\frac{1}{2}\sum_{\gm=1}^{m-1}c^\dg_{i\gm\sg}\boldsymbol{\tau}^\pdg_{\sg\sg'}c^\pdg_{i\gm\sg'}$, $\boldsymbol{\tau}$ are the $2\times2$ Pauli matrices, and
$\ij$ indicates summation over nearest-neighbor lattice sites. 
In the strong-coupling limit for lower orbitals ($U\gg t_\gamma$) and $n_\gamma\approx1$, according to the Schrieffer-Wolff transformations the AFM coupling amplitude scales as ${\cal J}_{\rm A}\propto t_\gm^2/U$ ($\gm<m$) 
\footnote{Eq.~\eqref{extKLM} is valid for the system that is rotationally-invariant in spin space. Otherwise, at $\alpha\neq1$, the magnetic couplings become anisotropic, but their scaling behavior with $t$ and $U$ remains similar.}.

\begin{figure}
\includegraphics[width=\linewidth]{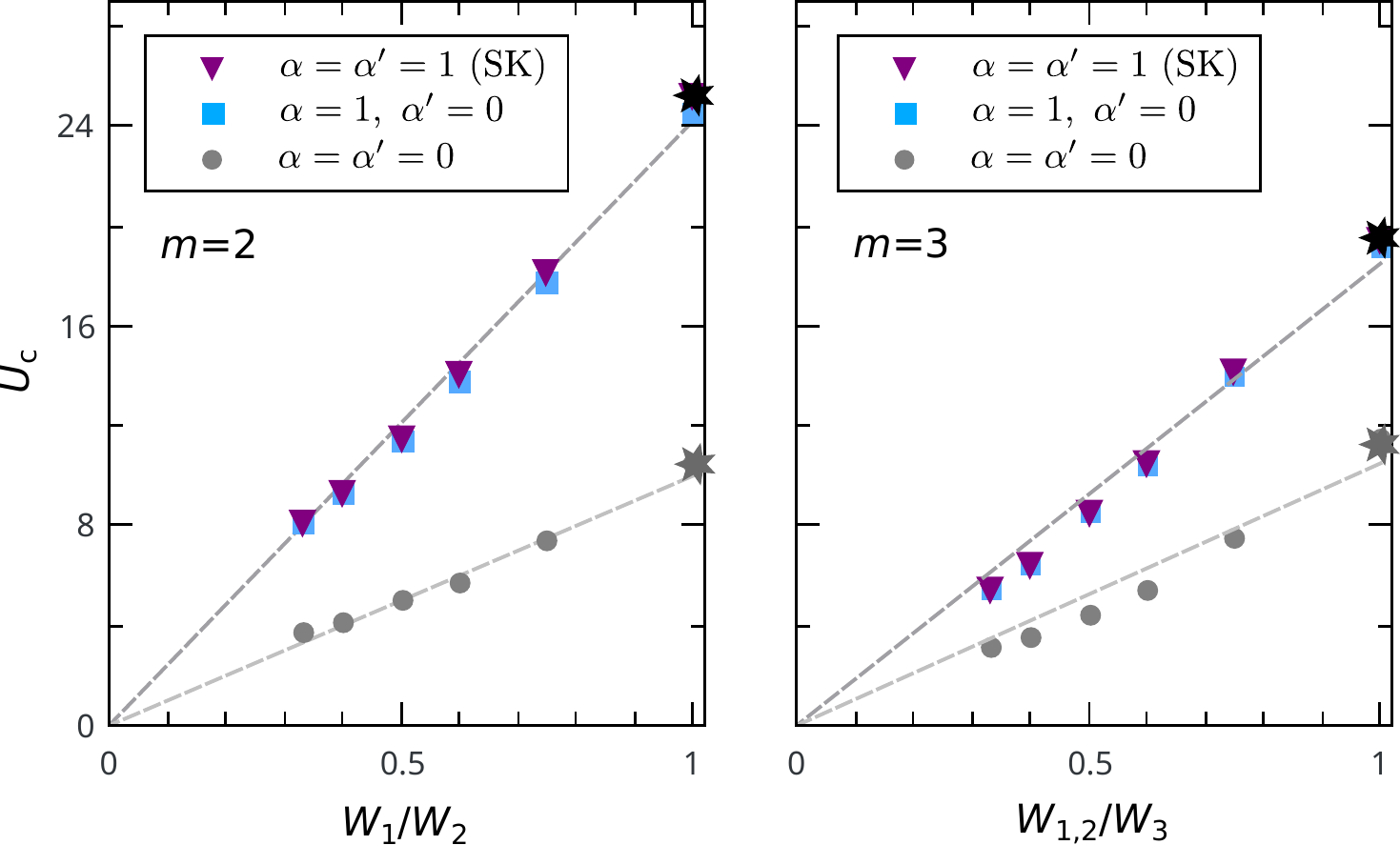}
    \caption{\label{fig4}
    Dependencies of the critical interaction strengths~$U_c$ for FM ordering on the bandwidth asymmetry in two-orbital (left) and three-orbital (right) Hubbard models at $\Delta=1.2$,
    $J=U/4$, $n=m-1+0.2$, and $T=0.025$. The star-shaped points at equal bandwidths correspond to the same points indicated in the phase diagrams in Fig.~\ref{fig3}. The linear fits (dashed lines) are based on the balance between the effective FM and AFM couplings in Eq.~\eqref{extKLM}; see text. The hopping amplitude in the upper band is fixed to $t_m=1$.}
\end{figure}

To show the importance of the AFM exchange processes that are explicitly included in the present extension of the DE model~\eqref{extKLM}, we perform an additional analysis, where the crystal field is kept fixed ($\Delta=1.2$), while the bandwidths~$W_\gm$ ($\gm<m$) are changed. Due to kinetic exchange mechanism, this is sufficient to suppress the AFM coupling ${\cal J}_{\rm A}$. This is supported by the DMFT results in Fig.~\ref{fig4}, which exhibit an enhancement of the FM instability due to suppression of its main competitor, the AFM coupling, with the band asymmetry.
As before, from direct comparison of $\alpha=0$ and $\alpha=1$ parametrizations we observe a more pronounced effect of SU(2) spin symmetry in the two-orbital model ($m=2$) that decreases with inclusion of an additional orbital flavor into the lower effective band ($m=3$).

The behavior of the critical coupling~$U_c$ on the band asymmetry in Fig.~\ref{fig4} can be understood as follows. Keeping the itinerant band fixed, the AFM-FM transition at zero temperature and under restriction of collinear magnetic ordering is expected to happen at a particular value of ${\cal J}_{\rm A}/J$. Given that $J\propto U$ (the ratio $U/J$ is kept fixed) and ${\cal J}_{\rm A}\propto t_\gm^2/U$, we obtain the result $U_c\propto t_{\gamma}$. The corresponding behavior is indicated by fits in Fig.~\ref{fig4} that shows relatively good agreement with the DMFT data that are obtained by considering the full model~\eqref{hamilt} in the DE regime at nonzero temperature.

Figure~\ref{fig3} shows that the DE description is not valid at low $\Delta$. The stepwise changes in positions of the FM phase boundaries, 
according to the panels shown on the right-hand side of Fig.~\ref{fig3}, are directly related to the change from the regime of partially overlapping bands (with the metallic behavior of all orbital flavors) to the DE limit (with the metallic behavior of only $\gamma=m$ flavor). Note also that for the SU(2)-symmetric case and $m=3$ (in contrast to $m=2$) it is possible to drive the system directly from the FM to the AFM ordered state by changing only the magnitude of the splitting $\Delta$.

The regime of intermediate crystal fields is interesting in several aspects. First, the ferromagnetism with partially overlapping bands is realized in high-valence transition-metal oxides, such as SrCoO$_3$ \cite{Caciuffo99, Kunes12sco}. Second, this can be experimentally studied with ultracold gases of alkaline-earth atoms in optical lattices (see Refs.~\cite{Scazza2014NP, Zhang14S, Riegger18}), where the occupations of the split bands can be tuned independently, therefore, the magnitude of the splitting~$\Delta$ can be changed in a wide range.

To keep the study consistent, we do not extend our current description to larger degeneracies of orbitals, but the observed behavior allows us to make a useful extrapolation from the point of view of solid-state realizations. For systems characterized by the electronic $d$ orbitals in cubic perovskite crystal structures (and in manganites, in particular), the typical band asymmetry can be roughly estimated as $W_{e_g}/W_{t_{2g}}\approx 2$. Assuming triple degeneracy of the lower ($t_{2g}$) and double degeneracy of the higher ($e_g$) states and $n=3+\delta$, the corrections to characteristic critical values due to spin-flip processes are expected to be still noticeable but, presumably, will not exceed 30\% (this agrees, in particular, with the recent studies~\cite{hausoel17}). The analyzed dependencies also suggest that the antiferromagnetic correlations within the $t_{2g}$ orbitals play an important role in physics of manganites and related compounds, thus must be properly accounted for.

\section{Summary and Outlook}\label{sec.4}
We studied the influence of spin symmetry (in particular, presence of the spin-flip term) in the interaction part of the Hamiltonian on the FM instability in the multiorbital Hubbard model by means of DMFT. We observe strong effects of suppression of FM phases when accounting for full spin-rotational symmetry in the two-orbital systems (in contrast to weaker effects for AFM ordering \cite{Antipov2012PRB,Golubeva2017PRB}). By considering the three-orbital model, it is shown that these effects become weaker (i.e., FM ordering effectively revives) with an increase of the number of active orbitals that agrees well with arguments based on suppression of quantum fluctuations due to approaching the limit of classical spins.
The analysis was extended to the case of split orbitals, where the corresponding transition from the Stoner to the double-exchange regime of FM ordering is observed, but the suppression effect originating from inclusion of the spin-flip processes remains significant.

The applied approach was restricted to measurements of observables diagonal in spin and orbital indices. Therefore, a number of instabilities with more complex structure (e.g., possible canted AFM ordering in the DE regime) were not studied directly. In view of recent developments in DMFT schemes with corresponding extensions~\cite{Hoshino2016PRB, shinaoka17, geffroy18}, research directions aiming to obtain more detailed phase diagrams in the regimes under study look realistic.

The results for the two-band model are also important from the viewpoint of the ultracold-atom experiments focused on approaching the ferromagnetic Kondo-lattice regime in optical lattices. Preliminary estimates for different parametrizations of interactions, $U\ll J\lesssim U'$, i.e., different from the current study but closer to the experimentally accessible values~\cite{Scazza2014NP,Riegger18}, show that the influence of spin-flip terms remains crucial for a determination of critical parameters for FM ordering. It is also interesting to study the influence of quantum fluctuations in two-orbital models with SU($N$)-symmetric interactions, where magnetic ordering is expected to be suppressed with an increase of $N$ (number of pseudospin flavors), in contrast to the studied direction of SU(2)-symmetric interactions and increasing $m$ (number of active orbitals), where FM phases become stabilized with $m$.

\begin{acknowledgments}
The authors thank A.~Golubeva, A.~Hariki, and P.~van~Dongen  for fruitful discussions. We highly appreciate technical assistance  by A.~Hausoel, P.~Gunacker, and G.~Sangiovanni with the \textsc{w2dynamics} package. A.S. and J.K. acknowledge funding of this work from the European Research Council (ERC) under the European Union's Horizon 2020 research and innovation program (Grant Agreement No. 646807-EXMAG). A.C. acknowledges funding of this work by the National Science Centre (NCN, Poland) under Grant No. UMO-2017/24/C/ST3/00357. Access to computing and storage facilities provided by the Vienna Scientific Cluster (VSC) is greatly appreciated. The calculations were also performed at the Poznan Supercomputing and Networking Center (EAGLE cluster).
\end{acknowledgments}

\begin{appendix}
\section{Magnetization behavior in the low-temperature regime}\label{app.A}
\begin{figure}
\includegraphics[width=\linewidth]{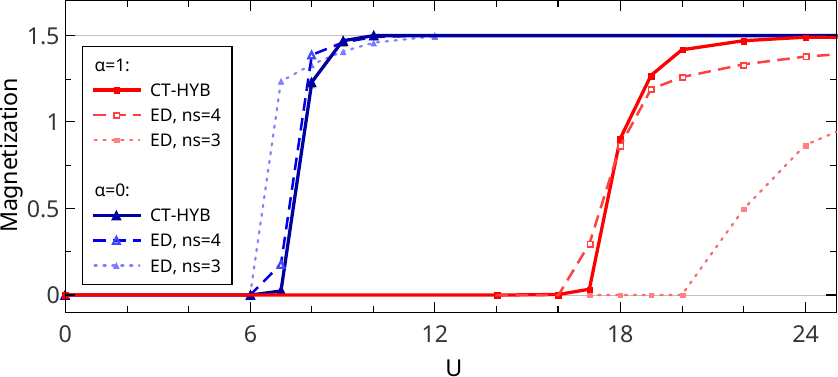}
    \caption{\label{fig6}
    Magnetization as a function of the interaction strength~$U$ for the two-orbital Hubbard model at $n=1.5$, $J=U/4$, and $T=0.025$. In the DMFT results obtained with the ED solver, $n_s$ denotes the number of effective bath orbitals per each orbital and spin flavor.}
\end{figure}
In Fig.~\ref{fig6} we provide an explicit comparison of the magnetizations $M=\sum_{\gm}(n_{\gm\ua}-n_{\gm\da})$ for two parametrizations ($\alpha=0$ and $\alpha=1$) obtained by DMFT with different (CT-HYB and ED) impurity solvers. Both parametrizations of interactions result in a fully polarized FM states in the limit of large interactions. Compared to more accurate data from the CT-HYB solver, the depicted dependencies also show the limitations of the ED solver due to the finite number of effective bath orbitals in the Anderson impurity model and indicate relatively good agreement at $n_s=4$.

At higher temperatures, the magnetization curves become smoothed out, i.e., the magnitude of the ordered moments decreases in the FM phases at the given values of $U$ and $J$. Qualitatively similar magnetization behavior is observed in the presence of the energy splitting~$\Delta$ between orbitals and in the three-orbital Hubbard model.
\end{appendix}

\bibliography{SU2_ferro}	
\end{document}